\documentclass[runninghead]{llncs}
\usepackage[T1]{fontenc}
\usepackage{svg}
\usepackage{pifont}
\usepackage{amsmath}
\usepackage{amssymb}
\usepackage{enumitem}
\usepackage{xspace}
\usepackage{stmaryrd} 
\usepackage{algorithm}
\usepackage{algpseudocodex}
\usepackage{multirow}
\usepackage{booktabs}
\usepackage{alltt}
\usepackage[subtle]{savetrees}

\usepackage[colorinlistoftodos]{todonotes}

\usepackage{subcaption}
\usepackage{pgf}
\usepackage{pgfplots}

\newcommand{\short}[1]{#1} %
\newcommand{\full}[1]{} %

\newcommand{\denotation}[1]{\ensuremath{\llbracket#1\rrbracket}\xspace}

\newcommand{\translate}[1]{\Box#1}

\newcommand{\etal}{\textit{et al.}\xspace}
\newcommand{\ie}{\textit{i.e.}\xspace}
\newcommand{\eg}{\textit{e.g.}\xspace}

\newcommand{\etc}{\textit{etc.}\xspace}

\newcommand{\system}{\textsc{Restricter}\xspace}
\newcommand{\cvc}{\textsc{cvc5}}

\newcommand{\mypara}[1]{\vspace*{0.55em}\noindent\textbf{#1}}

\newcommand{\rspace}[1]{\ensuremath{\mathcal{R}_{#1}}\xspace}
\newcommand{\alog}{\ensuremath{\mathcal{L}}\xspace}
\newcommand{\policy}{\ensuremath{\mathcal{P}}\xspace}
\newcommand{\policysub}[1]{\ensuremath{\mathcal{P}_{#1}}\xspace}
\newcommand{\prule}[1]{\ensuremath{R_{#1}}\xspace}
\newcommand{\rprule}{\prule{\star}^+}

\newcommand{\pred}[1]{\ensuremath{\mathsf{#1}}\xspace}
\newcommand{\schema}{\ensuremath{\Sigma}\xspace}
\newcommand{\environment}{\ensuremath{\mathcal{E}}\xspace}

\newcommand{\loc}{2100\xspace}
\newcommand{\proj}[1]{\ensuremath{\mathcal{L}\hspace{-0.3em}\downarrow_{#1}}\xspace}

\newcommand{\req}{\ensuremath{\texttt{req}}}

\usepackage{bbm}
\usepackage{wasysym}

\usepackage{listings}
\usepackage{xcolor}
\definecolor{c1}{HTML}{0033B3}
\definecolor{c2}{HTML}{1750EB}
\lstdefinelanguage{cedar}{
    keywords = [1]{ permit, when, deny, unless},
    keywords = [2]{ principal, action, resource},
    keywords = [3]{ &&, ||, has, ==, is, in}
    keywordstyle = [1]\color{c1},
    keywordstyle = [2]\color{c2},
    keywordstyle = [3]\color{orange},
    morestring = [b]",
  morestring = [b]',
  stringstyle = \color{purple}
}
\usepackage{graphicx}
\usepackage{hyperref}
\usepackage{color}

\usepackage{tcolorbox}

\begin{document}
\title{Automatically Tightening Access Control Policies with \system\thanks{This material is based on work supported in part by an Amazon Research Award and NSF award CCF-1954837. A full version of this paper is available at \url{https://arxiv.org/abs/XXXX.XXXXX}.}
}

\author{Ka Lok Wu\inst{1}
\orcidID{0000-0001-6315-9068}
\and
Christa Jenkins\inst{2}\thanks{Work done while the author was at Stony Brook University.}
\orcidID{0000-0002-5434-5018}
\and
Scott D. Stoller\inst{1}
\orcidID{0000-0002-8824-6835}
\and
Omar Chowdhury\inst{1}\thanks{Corresponding author.}\orcidID{0000-0002-1356-6279}}
\authorrunning{K. Wu et al.}
\institute{Stony Brook University, Stony Brook NY 11794, USA\\
\email{\{kalowu,stoller,omar\}@cs.stonybrook.edu}
\and
Galois, Inc. Portland, OR 97204, USA\\
\email{christa.jenkins@galois.com}}

\maketitle
\begin{abstract}
Robust access control is a cornerstone of secure software, systems, and networks. 
An access control mechanism 
is as effective as the policy 
it enforces. However, authoring effective policies that satisfy desired properties 
such as the \emph{principle of least privilege} is a challenging task even for 
 experienced administrators, as evidenced by many real instances of policy 
misconfiguration. 
In this paper, we set out to address this pain point by proposing 
\system, which \emph{automatically} tightens each (permit) policy rule of a  
policy with respect to an access log, which captures some 
already exercised access requests and their corresponding access decisions (\ie, allow or deny). \system achieves policy tightening by reducing the number of access 
requests permitted by a policy rule 
without sacrificing 
the functionality of the underlying system it is regulating. We implement 
\system for Amazon's Cedar policy language and demonstrate its effectiveness 
through two realistic case studies.
\keywords{Security Policy \and Access Control \and SyGuS.}
\end{abstract}
\section{Introduction}
\label{sect:introduction}
An access control mechanism is a critical defense 
for ensuring the resilient security posture of 
software, systems, and networks. In the   
2024 CWE Top 25 Most Dangerous Software Weaknesses 
published by MITRE \cite{mitre2024}, three of them were related 
to access control and authorization management\full{ 
(\ie, CWE-862: Missing Authorization, 
CWE-269: Improper Privilege Management, 
CWE-863: Incorrect Authorization)}. 
In the similar vein, 
OWASP's Top 10 list of critical web application security risks 
published in 2021 ranks ``\emph{Broken Access Control}'' at the 
top \cite{owasp2021}. One reason for such access control-related 
vulnerabilities can be attributed to  
access control checks being intertwined 
with application logic in source code, 
making it challenging for maintaining and understanding 
the access control being enforced. It is well 
understood that decoupling the access control checks 
from the application business logic ensures better 
management and maintainability. To promote such decoupling, 
many access control systems have been 
developed  (\eg, 
Casbin \cite{Casbin}, 
Open Policy Agent \cite{OpenPolicyAgent}, 
Amazon's Cedar \cite{CedarWeb}).

At the heart of an access control system is an \emph{access control 
policy}, which aims to precisely capture the 
conditions under which users may access critical resources. 
Designing such policies to prevent unauthorized access and allow  
legitimate users to carry out their business obligations is a challenging task\full{,
even for experts}.
Misconfiguring a policy can lead to the following 
pitfalls: \ding{202} overly permissive policies 
violate the \textit{principle of least privilege} (\ie, users possessing 
only the privileges needed to carry out their business tasks) and allow 
unauthorized users to access critical resources, opening the door to 
misuse; 
\ding{203} overly restrictive policies
prevent legitimate users from accessing resources necessary for 
performing their business tasks.\full{ Although policy 
misconfigurations of type \ding{203} often get 
rectified due to user complaining due to failure 
to carry out business operations, type \ding{202} misconfiguration 
can remain undetected, waiting to be exploited.} 
Examples of type \ding{202} policy misconfigurations abound in practice \cite{mitre2024}.
For instance, exploitable misconfiguration of type \ding{202} were identified
in mandatory access control policies written by experts for Android's SELinux \cite{chen2017analysis}.
\emph{This paper designs, develops, and evaluates \emph{\system}, an automated
  technique, based on Satisfiability Modulo Theory
  (SMT)~\cite{BSST21_Handbook-of-Satisfiability-SMT} and Syntax-guided Synthesis
  (SyGuS)~\cite{alur2018}, for reducing over-permissiveness in policies.}

Designing a policy that follows the principle of least privilege, which we call the \emph{tightest policy},
is a challenging and error-prone task\full{ since it requires the 
administrator to envision all possible business use cases and their required privileges, then write an 
access control policy enabling only those use cases}.
\full{In reality, policy authoring tends to be an  
incremental process, where a policy evolves based on the discovery of new business use cases unintentionally 
denied by the policy's previous iteration.}
Over-privileges (\ie, requests permitted by the policy but not needed for any current 
business use case) can easily creep in as policies evolve.  
Over-privileges can also result from imprecision in identifying the permissions required 
for a use case and imprecision in formulating policy rules that grant those permissions.
\system{}'s addresses over-privileges by \ding{172} inferring such 
unintended over-privileges permitted by the current policy and then \ding{173} 
refining (\emph{tightening}) the policy 
by removing them.

Besides scalability, the main technical challenge in realizing \system{}'s approach 
is identifying over-privileges (step \ding{172}).
Unfortunately, a precise characterization of over-privileges 
in the current policy is often unavailable in practice.
\system{}'s over-privilege inference is based on analysis of the current policy together with an \emph{access log} in which all attempted  access requests and their correct decisions are logged. Abstractly, an access log can be viewed as a sequence of $\langle \req, d\rangle$ pairs in which 
$\req$ is a concrete \emph{access request} (\ie, a request by a user to perform an action on a resource/object) and $d$ is the correct decision for $\req$ (\ie, grant/permit/allow or deny/forbid).
\system{}'s over-privilege inference is based on the insight that  
\emph{the over-privileges are a subset of the permissions granted by the 
current policy and not exercised in the log}.

\system focuses on Attribute-based Access Control (ABAC)~\cite{sandhu1998role}, 
which is widely used and more general than other access control paradigms such as 
role-based access control (RBAC).
ABAC policy tightening can be formulated 
as a \emph{program synthesis problem} and solved using a Syntax-Guided Synthesis (SyGuS) solver. 
Unfortunately, tightening an entire policy at once with SyGuS does not scale even for moderate-sized policies\full{ 
due to the inherent overhead of SyGuS}. \system addresses the scalability challenges by limiting 
the size of terms that are enumerated and by adopting a 
\emph{rule-level analysis}, tightening permit rules individually\full{ by strengthening
the conditions determining which requests are allowed by that rule}. 
Rule-level analysis is also amenable to parallelization.

\system is instantiated for a large subset of \href{https://www.cedarpolicy.com/}{Cedar} \cite{Cedar_OOPSLA2024}, 
an ABAC system developed by Amazon Web Services.
It is then evaluated on two case studies.
It is able to scalably infer the potential 
over-privileges and refine most (6 out of 8) of the deliberately loose permit rules by adding  
appropriate general conditions instead of point-solutions, while also removing
some over-privileges in the other cases. In summary,
this paper makes the following contributions:
\begin{enumerate}[nosep,noitemsep]
\item We formulate and formally define the access control policy tightening problem with respect to a given policy, access log, and policy state.
\item We present an approach that can incrementally tighten each policy rule using automated reasoning (\ie, SyGuS, SMT).
\item We empirically demonstrate \system{}'s effectiveness and scalability on two realistic case studies.
\end{enumerate} 

\section{Background on Cedar}
\label{sec:background}
\label{sec:cedar}

This section presents a primer on \short{Cedar}\full{ the Cedar access control policy
language}. %
Due to space restrictions, we assume readers have some familiarity with
ABAC, SyGuS, and SMT solvers. 
Further details on Cedar and these other topics are available from other sources
\cite{Cedar_OOPSLA2024,standard2013extensible,sygus,cegis,cegis2,SMTLIB_Haifa2010,SMTLibWeb}.
\short{Cedar~\cite{CedarWeb} is an open-source authorization policy language developed
(and used at scale, especially, to protect customer API end-points) by Amazon Web Services (AWS). }%
We chose Cedar as the target of \system to show that our
approach is applicable to an authorization policy language that is
practical, expressive, and actively used in industry,
but \system can also be extended to work on other ABAC policy languages.

\begin{figure}[t!]
\footnotesize
\begin{subfigure}{1.0\linewidth}
\begin{verbatim}
entity User { isPCChair: Bool, isPcMember: Bool};
entity Paper { authors: Set<User>, reviewers: Set<User>};
entity Review { ofPaper: Paper, author: User, isMetaReview: Bool };
action Read appliesTo {
principal: User, resource: [Review, Paper], context: {isReleased?: Bool}};
\end{verbatim}
  \caption{Conference management system schema (snippet)}
  \label{fig:background-cedar-example-hotcrp-schema}
\end{subfigure}
\begin{subfigure}{1.0\linewidth}
\begin{verbatim}
permit (principal, action == Action::"Read", resource is Paper)
when { principal.isPcMember };

permit (principal, action, resource is Review)
when { principal in resource.ofPaper.authors };
\end{verbatim}
  \caption{Conference management system policies (snippet)}
  \label{fig:background-cedar-example-hotcrp-policy}
\end{subfigure}
  \caption{Example Cedar schemas and policies}
  \label{fig:background-cedar-example}
  \vspace{-9px}
\end{figure}

\subsection{Cedar Primer}
\label{sec:cedar-primer}
We introduce some of Cedar's salient language features, using a conference
management system inspired by HotCRP as a motivating example.
This example comes from one of our case studies, which is described further in
Section~\ref{sec:eval-case-studies}\full{; the code listing in
Figure~\ref{fig:background-cedar-example} is adapted from this case study}.

\noindent\textbf{Schemas.} Cedar schemas define the data model for the
entities and actions to which policies apply.
Declarations of entity types specify the names  and types of the entities'
attributes.
Valid attribute types include Booleans, entity types, and sets.
Attributes can be \emph{mandatory} or \emph{optional}, indicated resp.\ by the
absence or presence of the suffix \texttt{\small ?} in their name.
As an example, 
in Figure~\ref{fig:background-cedar-example-hotcrp-schema}, the
entity type \texttt{\small Paper} declares that each paper has two
attributes: its sets of authors and reviewers.

Schemas also list the set of \emph{actions} and the types of entities to
which each action applies.
Optionally, action declarations may specify the type of contexts associated
with the action (more on this below).
For example, Figure~\ref{fig:background-cedar-example-hotcrp-schema} declares
the action \texttt{\small Read}, which a \texttt{\small User} may take on a
\texttt{\small Paper} or \texttt{\small Review}, and which comes with a context
containing an optional Boolean attribute indicating whether the resource has been
released.

\noindent\textbf{Policies.}
A Cedar policy is composed of \emph{rules}\footnote{In Cedar, these are called
``policy set'' and ``policy,'' resp. For our
presentation, we choose the terms more commonly used in access control literature. }.
Each rule is a Boolean-valued function of four parameters --- \texttt{\small
  principal}, \texttt{\small action}, \texttt{\small resource}, and
\texttt{\small context} (implicit) --- comprising:
\begin{itemize}[nosep]
\item an \textbf{\textit{effect}}, \texttt{\small permit} or \texttt{\small
    deny}, indicating whether to allow or prohibit requests to which the rule
  applies; %
\item a \textbf{\textit{scope}} constraining the principals, actions, and
  resources to which the rule applies; and

\item a \textbf{\textit{body}} (optional), which place further constraints on
  the parameters, usually involving their attributes.
\end{itemize}
\noindent\emph{Example:} In
Figure~\ref{fig:background-cedar-example-hotcrp-policy}, the second rule has
\textbf{\textit{effect}} \texttt{\small permit}, a \textbf{\textit{scope}} applying to
any \texttt{\small principal} and \texttt{\small action} but only \texttt{\small
resource}s of type \texttt{\small Review}, and a \textbf{\textit{body}} further
constraining the rule to only those \texttt{\small principal}s
which are authors of the paper the review concerns.

\noindent\textbf{Composing rule effects.}
When Cedar's authorization engine is given a request and policy,
conceptually it applies each rule in the policy to the request.
To assemble these results into a final decision, it resolves ambiguous cases
as follows.
\ding{182} \emph{Default Deny:} If no \texttt{\small permit} rule applies, the
request is denied.
\ding{183} \emph{Deny Overrides Permit:} If a \texttt{\small forbid} rule
applies, the request is always denied.

\noindent\textbf{Entity Hierarchies.}  %
Entity type declarations can also specify the types of \emph{parent} entities.
Each entity has zero or more parents.  The reflexive transitive closure of
the \emph{parent of} relation comprises Cedar's \emph{ancestor hierarchy}.
This hierarchy provides support for role-based access control: the
ancestor-descendant relationship can express membership of users in roles as
well as role hierarchy.

\section{Problem Definition} %
In this section, we present a motivating example and formally define the policy tightening problem that \system aims to solve. 

\subsection{Motivating Example}
\label{sec:motivation}
Suppose the program committee (PC) chairs of 
a popular computer security conference are setting up 
the HotCRP conference review system instance 
for this year. The current PC chairs obtained the policy 
from the previous year's chairs; this policy is the one from which
the listing in Figure~\ref{fig:background-cedar-example} samples.
However, the chairs anticipate 
receiving a substantial number of paper submissions this year, 
which induced the decision of partitioning the papers into mutually disjoint 
areas (\eg, usable security, software security, network security, \etc).
In addition, they decided to have a decentralized administration by 
assigning an area to each reviewer.
It is analogous to having a set of small conferences under 
a bigger conference. 

In preparation, they extended every \texttt{\small Paper} to have 
an \texttt{\small area} attribute of type \texttt{\small Area}, denoting the area assigned to a
paper.
The \texttt{\small User} type is updated to include a new Boolean attribute
\texttt{\small isAreaChair} and to replace the \texttt{\small isPcMember} attribute
with an optional attribute \texttt{\small pcMember} of type
\texttt{\small Area}, which when set indicates the user is a PC member for that area.
The first rule in the example is modified to reflect the change.
These modifications are shown in Figure~\ref{fig:modified_example}. 

\begin{figure}
\footnotesize
\begin{verbatim}
// modified schema (snippet)
entity User { isPCChair: Bool, isAreaChair: Bool, pcMember?: Area};
entity Area;
entity Paper { authors: Set<User>, reviewers: Set<User>, area: Area};
// modified policy (snippet)
permit (principal, action == Action::"Read", resource is Paper)
when { principal has pcMember };
\end{verbatim}
  \caption{Modified Cedar policy}
  \label{fig:modified_example}
\end{figure}

We may observe in the log that each reviewer accesses only papers in their area.
After simplification, \system is able to tighten the modified rule in Figure~\ref{fig:modified_example} to
{\footnotesize
\begin{verbatim}
permit(principal, action == Action::"Read", resource is Paper)
when { principal has pcMember && principal.pcMember == resource.area};}
\end{verbatim}
}
\noindent which exactly captures the semantics of our observation.

\subsection{Notations and Problem Definition}
\label{sec:motivation-prob-def}

\noindent\textbf{Notations.}
Let $\schema$ be the schema that specifies the entity types, the attributes associated with them, the allowed hierarchies, and the allowed principals and resources for each action. Let $\environment$ be the environment, also known as the entity store, that consists of the principals, resources, and their attributes.  
Let $\req=\langle \mathit{prin}, \mathit{act}, \mathit{obj}\rangle$ be an access request that consists of the principal, action, and resource, respectively.
Let \rspace{} be the universe of all requests %
consistent with the schema. 
Let $\pred{PolicyEval}(\req,\policy,\environment)$ be the function that evaluates a request $\req$ with respect to a policy $\policy$ and environment \environment according to Cedar's semantics, 
returning an access decision $d\in\{\pred{allowed},\pred{denied}\}$. 
Let \denotation{\policy} be the set of requests allowed by \policy, \ie, 
$\denotation{\policy}\overset{\Delta}{=}\{\req \mid  \req\in\rspace{} \wedge \pred{PolicyEval}(\req, \policysub{}, \environment)=\pred{allowed}\}$. If $\prule{}^+$ is one of the permit rules in $\policy{}$, then we can similarly define the set of requests allowed by $\prule{}^+$ as $\denotation{\prule{}^+}\overset{\Delta}{=}\{\req \mid  \req\in\rspace{} \wedge \pred{PolicyEval}(\req, \{\prule{}^+\}, \environment)=\pred{allowed}\}$.

An access log $\alog\subset \rspace{}\times \{\pred{allowed},\pred{denied}\}$ is
a set of requests labeled with the decision for that request from the
initial policy $\policy_{\text{init}}$.
Policy $\policy$ is \emph{consistent} with 
\alog, denoted $\policy \sim \alog$, 
if and only if 
$\forall \langle \req, d\rangle \in\alog. \pred{PolicyEval}(\req, \policy, \environment) = d$.
We 
assume 
$\policysub{\text{init}}\sim\alog$.

\mypara{The Policy Tightening Problem.} Given 
a type-safe Cedar policy $\policy_\mathsf{init}$, an access log \alog, a universe of 
access requests \rspace, an entity schema 
\schema, and an environment \environment, the \emph{policy tightening 
problem} is to synthesize a type-safe Cedar 
policy $\policy_\star$ such that: (I) $\policy_\star\sim\alog$,
and (II) $\denotation{\policy_\star} \subseteq \denotation{\policy_\mathsf{init}}$. 
The rationale for (II) is to avoid violating the least privilege principle. %
In what follows, 
``\emph{policy}'' refers to a type-safe 
Cedar policy unless mentioned otherwise. 
When not explicitly identified, we consider a fixed schema \schema 
and an environment \environment. 
We use $\policy_\star$ to denote the tightened policy throughout the paper.

It is easy to see that the policy-tightening problem is under-constrained, since any subset of the unexercised permissions can be removed.  
A trivial solution is to return $\policy_\mathsf{init}$;
our algorithm is designed to remove unexercised permissions
without changing the size of the policy too much and modifying the structure of the policy.

\section{Design Dimensions} %
\label{sec:designdimensions}
We discuss three design dimensions and their trade-offs that
one may consider in developing an approach like \system{}.

\mypara{Global vs. Local Tightening.} 
The first dimension is to consider whether to 
tighten the whole policy at once (\emph{global}) or 
tighten each policy rule individually (\emph{local}).
\begin{itemize}[noitemsep,nosep]
\item While costly, global tightening has the benefit of simplifying and
  minimizing the \emph{entire} policy during tightening.
\item While having a more limited view, local tightening can be more scalable:
  one can focus on the portions of \alog and entity store that are relevant to
  individual rules.
\end{itemize}
The local approach is further faciliated by Cedar's \emph{default deny, deny overrides permit}
semantics (Section~\ref{sec:cedar-primer}):
\emph{as permit rules are the only ones allowing accesses, it is sufficient to
  tighten only permit rules locally.} 
We can refine the policy tightening problem as follows.

Let the set of allowed log requests be $\alog^+\overset{\Delta}{=}\{\req| \langle \req, d\rangle\in\alog \wedge d=\pred{allowed}\}$, and the log slice with respect to permit rule $\prule{}^+$ be $\proj{\prule{}^+} \overset{\Delta}{=}
\alog^+\cap\denotation{\prule{}^+}$.
For policy tightening, it suffices to find a permit rule
$\prule{*}^+$ for each permit rule $\prule{}^+$ in the input policy such that:
\begin{enumerate}[nosep]
    \item[(I$'$):] $\rprule\sim \proj{\prule{}^+}\times \{\pred{allowed}\}$, and
    \item[(II$'$):] $\forall \req\in \rspace, \pred{PolicyEval}(\req,\{\prule{}^+\}, \environment) = \pred{denied}\implies\pred{PolicyEval}(\req,\{\rprule\}, \environment) = \pred{denied}$.
\end{enumerate}
Here, we need to focus only on the permitted part of the log in (I$'$) as the denied part is guaranteed to be consistent by (II$'$).

\mypara{Synthesizing vs. Incremental Strengthening.} 
The next design dimension is whether 
to generate a new policy or rule, replacing
the previous one, or strengthen the current policy or rule incrementally.
\begin{itemize}[noitemsep,nosep]
\item Synthesizing new policies or rules is \emph{prima facie} a more general
  approach, and has a larger search space.
\item Incremental strengthening\full{, being more limited,} has a much more tractable
  search space\full{ (our initial exploration confirms this hypothesis)}.
\end{itemize}

\emph{Permit vs. Deny.} Even when one chooses the latter approach, there are 
two  alternatives: (i) adding or broadening deny rules to remove
over-privileges, or (ii) strengthening permit rules to reduce over-privilege.
From the perspective of scalability, the search space for (ii) can be further
reduced as it suffices to identify additional conditions to be imposed on the
permit rules. The additional conditions can be in the form of restricting the
\textit{\textbf{scope}} or \textit{\textbf{body}} of the permit rule.
The conditions one can place in the \textit{\textbf{scope}} of a Cedar rule are
restricted to certain forms --- to restrictive for productively tightening.
For the \textit{\textbf{body}}, one can always add a
conjunctive formula (any Boolean-valued Cedar expression) to a rule to tighten
it.
This has the benefit of being amenable to incremental tightening of a rule by
adding one conjunct at a time.
\full{The resulting rule is also \emph{syntactically} guaranteed to be more
restrictive than the initial rule.}

\mypara{Encoding Approaches.} 
Irrespective of 
global/local, and synthesis/refinement approaches to tightening, a primary
challenge is to ensure that any automated reasoning approaches for policy
tightening not only scale with the size of the input but also generate
a tightened Cedar policy that is both type-safe and satisfies the restrictions
(I) and (II) (or their primed versions) discussed above.
\begin{itemize}[noitemsep,nosep]
\item We can take a \emph{syntactic approach} to policy encoding, modeling the
  abstract syntax tree (AST) of our supported subset of Cedar as an algebraic
  datatype (ADT); or
\item We can take a \emph{semantic approach}, translating the policy as a
  quantifier-free first-order logic (QF-FOL) formula.
\end{itemize}

The \emph{syntactic approach} (a.k.a. \emph{deep embedding}) is natural: the
policy to be synthesized is a constant of the ADT type, and $\pred{PolicyEval}$
is a recursive function over values of the ADT expressing Cedar's operational semantics.
One can then invoke the SMT solver's finite model finding capability for policy
tightening.
Unfortunately, in our evaluation such an approach suffers from severe
scalability issues, as \emph{user-defined ADTs are not amenable to optimizations
  enjoyed by native SMT theories.}

In the \emph{semantic approach} (a.k.a a \emph{shallow embedding})
a Cedar policy $\policy$ with a set of permit rules
$\{\prule{1}^+,\prule{2}^+,\ldots,\prule{m}^+\}$ and a set of forbid rules
$\{\prule{1}^-,\prule{2}^-,\ldots,\prule{n}^-\}$, can be viewed as a QF-FOL
formula: $\bigwedge_{i=1}^n \boxminus(\prule{i}^-) \wedge \bigvee_{j=1}^m
\boxplus(\prule{j}^+)$. Here, $\boxplus(\cdot)$ and $\boxminus(\cdot)$ are
functions that take a permit and deny policy rule, respectively, and return
their QF-FOL representation.
A rule has the form $\langle p, a, o, \phi\rangle $, where $p,a,o$ represent conditions on principals, actions, and resources respectively, and $\phi$ represents the body. %
A permit rule $\prule{}^+ = \langle p, a, o, \phi\rangle$ is
translated to
$\translate{p}\wedge \translate{a}\wedge\translate{o}\wedge \translate{\phi}$, where  
$\translate{c}$ is the translation (as a conjunctive formula in the QF-FOL fragment)
of a Cedar condition $c$.
Similarly, a forbid policy rule $\prule{}^- = \langle p, a, o, \phi\rangle$ is translated to
$\translate{p}\wedge \translate{a}\wedge\translate{o}\rightarrow \neg\translate{\phi}$ (notice the negation).

Therefore, we can represent a Cedar policy as an SMT term mimicking the above QF-FOL form and 
restricting ourselves mostly to native theories when possible.
$\pred{PolicyEval}$ conceptually reduces to the evaluation of an SMT term.
Finding a policy/rule becomes an instance of a SyGuS problem where the
function to be synthesized is 
a QF-FOL formula that captures the
evaluation of all the policies. We can also encode the original policy
rules as SMT terms and use them as semantic constraints. A challenge still remains: 
\emph{capturing Cedar type-safety as semantic constraints does not scale}.

\section{\system{}'s Approach}
\label{sect:methods}
\begin{figure*}[t!]\captionsetup[subfigure]{font=scriptsize}
{
    \begin{subfigure}[t]{0.21\textwidth}
\scalebox{0.5}{\begin{tikzpicture}
\fill[gray!35!white, even odd rule]
  (3,-2) rectangle (-3,2);
\fill[green!30!white] (2, -1.5) rectangle (-2, 1.4);
\fill[green!60!white] (-0.5, -1.5) rectangle (-2, 1.4);
\draw[black] (2, -1.5) rectangle (-2, 1.4);
\draw[black] (-0.5, -1.5) rectangle (-2, 1.4);

\node at (-1, 1.7) (P) {$\denotation{\prule{}^+}$};

\node[align=center] at (-1.25, 0) (L) {Log\\Slice};
\node[align=center] at (0.7, 0) (po) {Potential\\Over-privilege};
\end{tikzpicture}}
\subcaption{$\denotation{\prule{}^+}$ is partitioned into 
the log slice and potential over-privilege.}
\label{fig:ideaa}
    \end{subfigure}\hspace{1.5em}
    \begin{subfigure}[t]{0.21\textwidth}
\scalebox{0.5}{\begin{tikzpicture}
\fill[gray!35!white, even odd rule]
  (3,-2) rectangle (-3,2);
\fill[green!30!white] (2, -1.5) rectangle (-2, 1.4);
\fill[green!60!white] (-0.5, -1.5) rectangle (-2, 1.4);
\draw[black] (2, -1.5) rectangle (-2, 1.4);
\draw[black] (-0.5, -1.5) rectangle (-2, 1.4);

\node at (-1, 1.7) (P) {$\denotation{\prule{}^+}$};

\node[align=center] at (-1.25, 0) (L) {Log\\Slice};
\node[draw,circle, fill=black, inner sep=0pt, minimum width=0.5pt] (r_pt) at (1, 1) {};
\node at (1.2, 1.2) (r) {$\req$};
\node[align=center] at (0.7, 0) (po) {Potential\\Over-privilege};
\end{tikzpicture}}
\subcaption{Choose a request $\req$ in the set of potential over-privileges.}
\label{fig:ideab}
    \end{subfigure}\hspace{1.5em}%
     \begin{subfigure}[t]{0.21\textwidth}
\scalebox{0.5}{\begin{tikzpicture}
\fill[gray!35!white, even odd rule]
  (3,-2) rectangle (-3,2);
\fill[green!30!white] (2, -1.5) rectangle (-2, 1.4);
\fill[green!60!white] (-0.5, -1.5) rectangle (-2, 1.4);
\fill[red!35!white] (1.5, -2) -- (3, -2) -- (3, 2) -- (0.7, 2);
\draw[black] (2, -1.5) rectangle (-2, 1.4);
\draw[black] (-0.5, -1.5) rectangle (-2, 1.4);

\node at (2.5, 1.5) (phi) {$\denotation{\neg \phi^*}$};

\node at (-1, 1.7) (P) {$\denotation{\prule{}^+}$};
\node[align=center] at (-1.25, 0) (L) {Log\\Slice};
\node[draw,circle, fill=black, inner sep=0pt, minimum width=0.5pt] (r_pt) at (1, 1) {};
\node at (1.2, 1.2) (r) {$\req$};
\node[align=center] at (0.7, 0) (po) {Potential\\Over-privilege};
\end{tikzpicture}}
\subcaption{Use SyGuS to synthesize a predicate $\phi^*$ that denies $\req$.}
\label{fig:ideac}
    \end{subfigure}\hspace{1.5em}
     \begin{subfigure}[t]{0.21\textwidth}
\scalebox{0.5}{\begin{tikzpicture}
\fill[gray!35!white, even odd rule]
  (3,-2) rectangle (-3,2);
\fill[green!30!white] (2, -1.5) rectangle (-2, 1.4);
\fill[green!60!white] (-0.5, -1.5) rectangle (-2, 1.4);
\fill[gray!35!white] (1.5, -2) -- (3, -2) -- (3, 2) -- (0.7, 2);
\draw[black] (-0.5, -1.5) rectangle (-2, 1.4);
\draw[black] (-2, -1.5) -- (-2, 1.4) -- (0.82, 1.4) -- (1.4, -1.5) -- (-2, -1.5);

\node at (-1, 1.7) (P) {$\denotation{\rprule{}} = \denotation{\prule{}^+}\cap\denotation{\phi^*}$};
\node[align=center] at (-1.25, 0) (L) {Log\\Slice};
\node[draw,circle, fill=black, inner sep=0pt, minimum width=0.5pt] (r_pt) at (1, 1) {};
\node at (1.2, 1.2) (r) {$\req$};
\node[align=center] at (0.31, 0) (po) {\scriptsize Potential\\\scriptsize Over-privilege};
\end{tikzpicture}}
\subcaption{The rule $\rprule{}$ that includes $\phi^*$ as a conjunct denies $\req$.}
\label{fig:idead}
    \end{subfigure}
    }
    \caption{An illustration of \system{}'s main idea for Step \ding{203}. \system chooses requests in the set of potential over-privileges
    to be denied by a conjunct $\phi$ (to be synthesized with SyGuS) while keeping the log slices permitted.}
    \label{fig:idea}
\end{figure*}
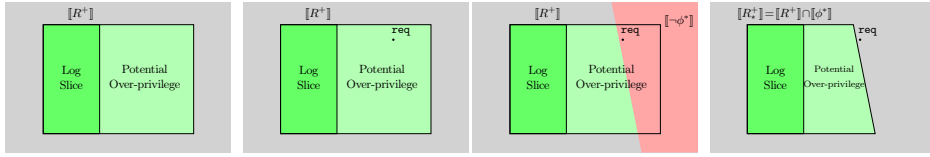

\newcommand{\initpolicy}{\ensuremath{\policy_\mathsf{init}}\xspace}
\newcommand{\tightenedpolicy}{\ensuremath{\policy_\mathsf{\star}}\xspace}

\system 
 takes as input the current policy 
$\policy_\mathsf{init}$, an environment $\environment$, and
an access log $\alog$, and generates a tightened policy $\policy_\star$.
Out of the design choices mentioned 
in Section~\ref{sec:designdimensions}, \system{}'s approach embraces the following choices: 
(a) local, rule-based tightening; 
(b) incremental strengthening of each permit rule of the input policy through the introduction of conjunctive conditions; and
(c) SyGuS-based approach for identifying the conjunctive conditions for rule strengthening. 
\emph{By adding conjunctive formulas to permit rules, \system{} syntactically guarantees that the tightened policy $\policy_\star$
  satisfies $\denotation{\policy_\star} \subseteq
  \denotation{\policy_\mathsf{init}}$}.

\noindent\textbf{Sketch of \system{}'s Approach.}
\system's algorithm is presented in Algorithm~\ref{main_algo}. At a high level, 
\system{}'s approach is as follows: \ding{202} collect each permit rule $\prule{i}^{+}$ from 
\initpolicy; \ding{203} iteratively strengthen each permit rule $\prule{i}^{+}$ from \initpolicy \emph{in parallel} 
to obtain a tighter permit rule $\prule{i\star}^+$; and \ding{204} collect all the forbid rules of \initpolicy 
and combine them with the newly strengthened permit rules $\prule{i\star}^+$ to obtain the tightened policy 
\tightenedpolicy. In the rest of the section, we focus mainly on Step \ding{203} of the overall approach. 

\noindent\textbf{High-level View of Tightening a Permit Rule (Step \ding{203}).}
Before delving into the details, we first present a high-level view of \system{}'s approach 
to Step \ding{203}; an illustration of the ideas is also presented in Figure \ref{fig:idea}. The rest of this section expands upon each of the following 
steps. \system iteratively performs the following steps, while 
using the obtained strengthened policy of an iteration as input to 
the next,
until some termination condition is met. 
\begin{description}[nosep]
    \item[i. Calculate Relevant Log Slice:] 
    We first calculate \proj{\prule{i}^{+}}, the set of permitted requests in \alog that are also 
    permitted by $\prule{i}^{+}$.
    \item[ii. Calculate Approximate Over-Privileges in $\prule{i}^{+}$:] We then
      calculate an over-approximation of the set of over-privilege requests $\mathbb{R}$,
      such that all $\req\in\mathbb{R}$ satisfies $\req\not\in\mathsf{dom}(\alog)$
      (\(\req\) is not in the log) and $\req$ is permitted by $\prule{i}^{+}$.
    \item[iii. Calculate Type-safe Cedar Predicate List:] We precalculate the candidate predicates that SyGuS can consider during the term enumeration phase. 
    This approach avoids
    needing to encode Cedar's type checking rules into SyGuS, and instead delegate it to the meta-program generating the SyGuS problem instance. %
    \item[iv. Invoke SyGuS:] We finally invoke SyGuS to generate a type-safe Cedar predicate $p$ such that 
    $\boxplus{(\prule{i}^{+})} \wedge \translate{(p)}$ (\ie, the
    tighter rule $\prule{i\star}^+$) allows every request in $\proj{\prule{i}^{+}}$, and denies at least one request in $\mathbb{R}$.
\end{description}

\subsection{Synthesizing a tighter permit rule (Step \ding{203}).}
\label{sec:mainalgo}

\begin{algorithm}[!t]
\caption{Main algorithm of \system}\label{main_algo}
\begin{algorithmic}[1]
\Procedure{Restrict}{$\policy_\mathsf{init}$,\alog, \rspace, \schema, \environment, $t$}
\State $\policy_*\gets\emptyset$
\ForAll{positive rule $\prule{}^+$ of $\policy_\mathsf{init}$}
\State $\rprule\gets \prule{}^+$
\Repeat
\If{$\denotation{\rprule}\setminus\proj{\prule{}^+}=\emptyset$}
{\bf break} \EndIf
\State $\pred{POP}\gets \denotation{\rprule }\setminus\proj{\prule{}^+}$\Comment{Steps i. and ii.}
\State Pick requests ${\{\req_i\}}_i \subset \pred{POP}$ %
    \State $\rprule\gets $ \textsc{Restrict\_one}($\rprule$, $\proj{\prule{}^+}$, \schema, \environment, ${\{\req_i\}}_i$) \Comment{Steps iii. and iv.}
    \If{$\rprule=\bot$} restore the previous $\rprule$ \EndIf
\Until{$\rprule=\bot$ for $t$ times}
\State Insert the previous value of $\rprule$ into $\policy_*$
\EndFor
\State Insert all negative rules $\prule{}^-$ into $\policy_*$
\State \textbf{return} $\policy_*$
\EndProcedure
\end{algorithmic}
\end{algorithm}

We now present descriptions of Steps (i) - (iii).
Details of Step (iv) are in Section \ref{sec:implementation}. 

\mypara{Finding a Separating Predicate for  Tightening.} Figure~\ref{fig:idea}
shows the main idea of tightening a permit rule.
We over-approximate the over-privileges as the accesses that do not appear in the
log. Given the log and the permit rule, we compute the slice
$\proj{\prule{}^+}$. Then, the set of \emph{potential unexercised
  over-privileges} $\pred{POP}\overset{\Delta}{=} \denotation{R^+} \setminus
\proj{R^+}$ consists of the requests permitted by the rule but not in
$\proj{\prule{}^+}$(Figure~\ref{fig:ideaa}). 

Given a permit rule $\prule{}^+=\langle p, a, o, \phi\rangle$, 
we use 
SyGuS to find a \emph{separating} predicate $\phi^*$ such that $\rprule{}=\langle p, a, o, \phi \wedge \phi^*\rangle$ is consistent with the log slice, and denies some requests in \pred{POP} (Figure~\ref{fig:ideac}), 
Here, the restrictiveness requirement (II') can be dropped because $\rprule$ is syntactically guaranteed to be more restrictive. Denying some requests in \pred{POP} makes $\rprule{}$ \emph{strictly} more restrictive than $\prule{}^+$.
We curate a list of candidate predicates that are type-safe, and the syntactic constraint becomes choosing one of the candidate predicates.
If there is a predicate that satisfies the constraints, then $\rprule{}$ is the solution. Otherwise, $\bot$ is returned.
\mypara{Avoiding Existential Quantification.}
To formulate that the separating predicate denies \emph{some} (unspecified)
request, we would need existential quantification.
To avoid this, we explicitly choose some requests $\{\req_i\}_i$ in \pred{POP}, and assert that at least one of them 
is denied in $\rprule{}$ (Figure~\ref{fig:ideab}).
The new rule $\rprule{}$ (if SyGuS returns one) is strictly more restrictive than $\prule{}^+$ (Figure~\ref{fig:idead}).
We can repeat this process until either \pred{POP} is empty, or we fail to
generate a predicate for a specified number of iterations (with different chosen requests $\req_i$). The former
means that we have obtained the tightest rule that only permits everything in
the $\proj{R^+}$.
Section~\ref{sect:discussion} discusses 
other termination conditions.

\mypara{Generating Type-safe Predicates.}
To generate type-safe predicates, we analyze the Cedar schema and policy in the meta-program that generates candidate predicates. Then, we enumerate relevant type-safe predicates (Section~\ref{sect:symbolic-compilation}) 
in our own symbolic encoding to be fed to SyGuS as syntactic constraints. 

\subsection{Choosing Concrete Requests to Deny}
\label{sect:pointpicking}
\mypara{Explicit Enumeration.} We can choose a random point in $\pred{POP}$, or try to restrict every point in $\pred{POP}$ in parallel (see Figure \ref{fig:ideab}).
Both would require explicit computation of the set $\pred{POP}$, which entails enumerating over the set of requests $\rspace{}$ (with types restricted when possible) and checking whether $R^+_*$ applies. As the size of $\rspace{}$
grows with the number of entities, this would introduce a large  overhead.

\mypara{Request Generation with SMT Solver.} Alternatively, we can take the SMT
encoding of the permit rule $\prule{}^+$, and ask the SMT solver to generate a concrete request
$\req$ that satisfies $\prule{}^+$ and is not in 
$\proj{R^+}$.
If SyGuS fails to produce a tightened rule, then we add the picked request to the
set of requests to be blocked and repeat. 
To reduce the likelihood of the SMT solver returning similar requests in different iterations, we randomize the entity encoding in our SMT representation, specifically, the mapping from entities to integers (discussed 
 in Section~\ref{sect:symbolic-compilation}). 
\begin{algorithm}[!t]
\caption{Restrict a rule by adding one conjunct}\label{restrict-one}
\begin{algorithmic}[1]
\Procedure{Restrict\_one}{$\prule{}^+$, $\pred{Log\_Slice}$, \schema, \environment, $D$}
\State $\langle p, a, o, \phi\rangle\gets$ symbolic encoding of $\prule{}^+$ under $\Sigma$
\State $P\gets$ Set of type-safe candidate predicates\Comment{Step iii.}
\State Ask SyGuS to construct a function $f: \rspace{}\to \{\pred{allowed},\pred{denied}\}$ that takes the form of a rule $\langle p, a, o, \phi\wedge \phi^*\rangle$ where $\phi^*\in P$, such that \begin{enumerate}
    \item $\forall \req\in \pred{Log\_Slice}, f(\req)= \pred{allowed}$, and
    \item $\bigvee\limits_{\req^*\in D} \left(f(\req^*)=\pred{denied}\right)$ holds\Comment{Step iv.}
\end{enumerate}
\If{SyGuS fails} \textbf{return} $\bot$ \EndIf
\State $\prule{\star}^+\gets$ encoding of $\langle p, a, o, \phi\wedge \phi^*\rangle$ in Cedar

\State \textbf{return} $\prule{\star}^+$
\EndProcedure
\end{algorithmic}
\end{algorithm}

\section{Implementation}
\label{sec:implementation}

\full{We now discuss \system{}'s implementation, including encoding the policy tightening problem as a SyGuS problem.}
\system is implemented in $\sim\!\!\loc$ lines of Python and $\sim\!\!400$ lines of Rust. 
The main challenge is to implement the function \textsc{Restrict\_one} (Algorithm~\ref{restrict-one}). We first encode the input Cedar policy using our symbolic compiler as an SMT term.
Along with the encoded policy, entity store and problem constraints, we invoke
\cvc{}'s~\cite{BBBKLMMMNOPRSTZ22_CVC5-Industrial-Strength-SMT} SyGuS
engine~\cite{RBNBT19_CVC-SyGuS}, which takes semantic constraints and syntactic constraints in the form of SMT terms, to solve the tightening problem.
The resulting rule, encoded as an SMT term, is translated back to a Cedar rule
as output. The request generation routine also uses a fragment of the symbolic compiler to encode the policy rules.

\subsection{Symbolic Encoding of Cedar}
\label{sect:symbolic-compilation}
We now discuss our encoding of entity stores and policies as SMT terms 
to make them amenable to constraint solving (See Section \ref{sect:pointpicking}) and synthesis (Finding separating predicate as discussed in Section \ref{sec:mainalgo}). We also discuss 
how our approach differs from that of Cutler \etal's
symbolic compiler~\cite[Section 4]{Cedar_OOPSLA2024}.

There are three main ways that our encoding differs from the one in Cutler \etal
\emph{First}, Cutler \etal's encoding does not consider specific entity stores,
so operators like hierarchy relations are left as uninterpreted functions.
We are given a concrete entity store, and can encode the hierarchy with concrete
values.

\emph{Second}, Cutler \etal's \emph{encoder} treat different entities as different types. The encoder meta-program instantiates the concrete types and pre-evaluate some expressions into the SMT solver.  In contrast, the policy evaluation function in the \emph{SMT solver} should take all possible entity types that are permitted by the schema as possible inputs. Therefore, we define one general entity type in the encoder and allow entities of any type to act as input to our function to synthesize, then make sure we only produce cedar-type-safe predicates in the encoder.
As an example, consider the expression ``\texttt{resource has ofPaper}'' using the schema in~\ref{fig:background-cedar-example-hotcrp-schema} that is true when the type of \texttt{resource} is \texttt{Review}, and false otherwise. It is not possible to reason on the types of SMT terms using SMT terms. The approach in Cutler \etal tries each instantiation of types for \texttt{resource}. For type \texttt{Review}, for example, the expression is compiled directly to the SMT term \texttt{true}. In contrast, we have to evaluate this expression inside the SMT solver, since we cannot determine the concrete type of \texttt{resource} that goes into the policy evaluation beforehand. Therefore, our encoding of \texttt{has} would be a function that takes the entity type and consists of \texttt{ite}'s that test on the \texttt{type} (which is an element in the constructor of the one general entity type) and the name of the attribute.

\emph{Finally}, we try to use simpler theories when possible in our encoding. As an example, we encode the name of the entities as 
integers instead of strings as the entity store only contains a small number of entities, 
and we can avoid invoking the string solver that may be expensive.

\mypara{General entity type.} 
In Cedar, an entity has a type and a name string.
In our approach, all entities are encoded using a single datatype $E$ that has a single constructor with \texttt{type} and \texttt{name} as integer arguments instead of the string type used in Cedar. We 
map each type in the schema and entity name in the entity store to an integer. In the semantic constraint, we state that the range of type and id of principals and resources is bounded by the image of these mappings. Equalities over entities are simply equalities over the type and id in the constructors. Representing these as integers rather than finite strings makes reasoning more efficient.

\mypara{Entity attributes.} The attributes of each entity type are defined in the schema with the name $f$ and type $t$. 
For each attribute name-type pair $f: t$ in the entity store, we define a function $\mathit{get}_{f, t}: E \to \text{Option}\ t$ that takes an entity of any type that are in the entity store and returns either the value of the attribute of the entity or \texttt{None} when either (1) the attribute does not exist for the type, or (2) the attribute is marked optional for the entity in the store, and the given entity does not have the attribute. We also have for every attribute a predicate $\mathit{has}_{f}: E\to \text{Bool}$ that returns \emph{true} if $f: t$ is an attribute for $e$ in the entity store for some type $t$.

\mypara{Entity Hierarchy.} We encode the hierarchy relation of the given entity store by pre-computing the reflexive transitive closure of the relation as a concrete predicate.

\subsection{Syntactic Constraints}

We now describe  the syntactic constraints that are posed to SyGuS to restrict
candidate rules.
If the original rule is $\langle p, a, o, \phi\rangle$, then the output rule has
the form $\langle p, a, o, \phi\wedge \phi^*\rangle$, where $\phi^*$ is a
type-safe Cedar predicate.
We pre-compute the list of candidate type-safe Cedar
predicates.
Table~\ref{tab:cedar-support} shows the subset of Cedar that \system can parse,
as well as the degree of \system{}'s ability to generate predicates utilizing
the listed features.
The features with only partial support for predicate generation, such as set
membership tests, are intentionally limited to avoid enumerating arbitrary constants and
instead only search applicable combinations of parameters, their attributes, and constants that appear in the entity store.
Features not listed in Table~\ref{tab:cedar-support}, such as Cedar's various
extension types, cannot be consumed by the current version of \system{}.

\begin{table*}[t]
  \centering
  \begin{tabular}{rcl}
    Feature & Support & Examples
    \\\hline \hline
    Equality & \CIRCLE & \texttt{principal.role == Role::"Teacher"}
    \\ Inequality & \CIRCLE & \texttt{User::"Alice" != User::"Bob"}
    \\ Attribute presence & \CIRCLE & \texttt{principal has pcMember}
    \\ Entity type test & \CIRCLE & \texttt{principal is Student}
    \\ Boolean negation & \CIRCLE & \texttt{!(principal in resource.authors)}
    \\ \hline
    Integer comparison & \LEFTcircle & \texttt{principal.balance >= resource.cost}
    \\ Hierarchy membership & \LEFTcircle & \texttt{principal in resource.course}
    \\ Set membership & \LEFTcircle & \texttt{principal in resources.authors}
    \\ \hline
    Conjunction, Disjunction & \Circle & \texttt{true \&\& true} \quad \texttt{true || true}
    \\ Integer operators & \Circle & \texttt{4 * 10 + 5 - 3}
    \end{tabular}

    {\footnotesize
      \CIRCLE: Full predicate generation support
      \quad \LEFTcircle: Limited predicate generation support\\
      \quad \Circle: No predicate generation support }

    \caption{Summary of \system{}'s support for Cedar.}
    \label{tab:cedar-support}
\end{table*}

\mypara{Type-safety.}
Due to the difference between our encoding
and that of Cutler \etal in
representing types, we need to analyze the typing requirements outside SyGuS to
ensure type-safety.
For example, the rule $\mathsf{ permit(principal, action, resource)}$ $\mathsf{when\mbox{ } \{principal.isAdmin\};}$
is not type-safe in Cedar
unless all entity types that can be principals have the Boolean attribute \verb|isAdmin|.
On the other hand, the rule 
{$\mathsf{ permit(principal, action, resource)\mbox{ }when\mbox{ }\{\mathbf{principal\,is\,User}}$ $\&\&\mathsf{\ principal.isAdmin\};}$}
is type-safe if \verb|User| has the Boolean attribute \verb|isAdmin|. To ensure our predicates are type-safe in Cedar, 
we syntactically restrict the types that appear in the expressions.

\mypara{Equality.} Since equality requires both sides to have the same type, we
can enumerate possible equalities by enumerating all constants of each type, identifying entities with attributes of that type, and generating equalities between them. Enumerating the constants is straightforward, since the environment is known, and we can collect the values that appear in attributes of known entities. The main challenge is identifying when an entity has an attribute of a given type. To ensure an entity $e$ has a given attribute $f$ of type $t$, 
we can add a guard $\mathit{has}_{f}(e)$ at the front.
Alternatively, for an entity $e$ of type $t_e$, if $t_e$ has a (required) attribute $f:
t$, we can add the guard {\tt $e$ is $t_e$}. %
Now, for every entity $e_1: t_1$ and entity $e_2: t_2$ that have attributes $f_1: t$ and
$f_2: t$, respectively, we can write the following type-safe predicate as a candidate predicate for \system: $e_1\text{ \tt is }t_1\wedge e_2\text{ \tt is }t_2\longrightarrow e_1.f_1 == e_2.f_2$. We also consider equality where one side is a constant like $e\text{ \tt is }t\longrightarrow e.f == const$ where $f$ and $const$ have the same type.

Similar constructions work for set membership and hierarchies. %
For example, for set membership, we can write the type-safe predicate for \system: $ e_1\text{ \tt is }t_1\wedge e_2\text{ \tt is }t_2\longrightarrow e_1.f_1 \in e_2.f_2$ if the type of $f_1$ is $t$ and type of $f_2$ is $\mathtt{Set}\ t$.

\mypara{Attribute Chains.} In Cedar, we can chain attributes together when the intermediate attributes have Entity or Record types. Using the schema in Figure~\ref{fig:background-cedar-example-hotcrp-schema}, if \texttt{resource} has type \texttt{Review}, we can write \texttt{resource.author.isPCChair}. It may be possible, depending on the schema, to have arbitrary nested or even cyclic attribute chains. For example, if \texttt{Paper} has a \texttt{metareview} attribute of type \texttt{Review}, then we can nest an arbitrary number of \texttt{.ofPaper.metareview} in \linebreak\texttt{resource.ofPaper.metareview}.

While generating candidate expressions, instead of leaving the attribute accesses recursive on the grammar level, we enumerate all type-safe attribute chains up to a specified depth, which is a parameter of our algorithm,
to limit our search space. We can ensure that the attribute chains are well-typed in Cedar, because we know the exact type of each attribute from the schema.
\section{Evaluation}
\label{sec:eval}
This section details our evaluation of \system, beginning with our guiding research
questions.

\mypara{Q1:} How frequently does \system produce desired tightenings?
When it fails to do so, how much over-privilege does it eliminate, and how much
legitimate privilege does it preserve?

\mypara{Q2:} How well does the performance of \system scale with respect to
entity store size and log size (\ie, the numbers of entities in the entity store and
access requests in the access log)?
 
Some challenges arise in answering these questions which further influence our
case study design and evaluations.

\noindent\textbf{Challenge 0:} Unfortunately, none of the previous works on policy tightening~\cite{dantoni2024automatically,eiers2023quantitative,mitani2023qualitative} open source the tools or the test cases they used for evaluation. As such, we are not able to benchmark against the previous works.

\noindent\textbf{Challenge 1:} 
Due to their sensitive nature, it is difficult
to obtain real-world examples of access control policies or logs.
\emph{Solution:} Our case studies are based on real-world management systems or examples
found in the access control literature~\cite{GClassroom,HotCRP}.

\noindent\textbf{Challenge 2:} Quantifying the answer to \textbf{Q1}
requires a precise characterization of a policy's over-privilege.
In practice, this is usually not available.
\emph{Solution:} Our case studies are designed with a loose initial policy and an intended tight policy. This intent is reflected indirectly in the access log
and used to evaluate the policy rules produced by \system, but is
never given to it directly.

\noindent\textbf{Challenge 3:} The sizes of the entity store and log are not the only factors
impacting the performance.  The shape of the entity store (\ie, the number and structure of relationships between entities) also affects performance.
\emph{Solution:} Our input generation approach is highly
flexible, allowing control over the shape as well as the size of the entity store.

\subsection{Design of Case Studies}
\label{sec:eval-case-studies}
We now overview the high-level design of our case studies, while the details can be found in the full version.
The two case studies used to evaluate \system are a classroom management
system inspired by Google Classroom~\cite{GClassroom}
and a conference management system
inspired by HotCRP~\cite{HotCRP}.
Figure~\ref{fig:eval-corresponding-pair} shows an example for the HotCRP case
study.
For each case study, we crafted a schema and two policies: 
\(\policy_{\mathsf{tight}}\), the desired tight policy; and
\(\policy_{\mathsf{init}}\), which we derived by deliberately introducing
over-privilege in some rules of \(\policy_{\mathsf{tight}}\) by dropping a conjunct. We generate the random entity stores and access logs for our case studies with a bespoke program (\(\sim\)3K LoC in Haskell). The program can generate sets of related entity stores and logs, which we dub ``families'', where entities and log entries in smaller members also appears in that of the larger members. The members are indexed by a \texttt{size} parameter. The size of the entity store and logs are roughly linear to the \texttt{size} parameter.

\begin{figure}[!t]
  \footnotesize
\begin{alltt}
// \(\policy\sb{\mathsf{tight}}\)
permit (principal, action in Action::"Read",resource is Paper)
when \char123 principal has isPcMember \&\& principal.pcMember == resource.area \char125;
// \(\policy\sb{\mathsf{init}}\)
permit (principal, action in Action::"Read",resource is Paper)
when \char123 principal has isPcMember \char125;
\end{alltt}%
  \caption{Example corresponding rule pair (HotCRP)}
  \label{fig:eval-corresponding-pair}
\end{figure}

\subsection{Evaluation Results}
\label{sec:eval-results}

We ran our experiments 
on a server with two Xeon Gold 5418Y CPUs totaling 96 threads and 256GB of RAM.
We used \cvc{} version 1.2.0 and the Cedar authorization engine version 3.2.1. For
each test case, we invoke \system to tighten every permit rule sequentially to
obtain accurate measurements, even though \system is parallelizable. 
We set the termination condition to 2 total failures, and generate three requests
as a potential over-privilege to be denied
in each
iteration.
For the datasets, we generated 100 dataset families using distinct seeds and varying log density from 30 to 100 percent with 10 percent increments.
We repeat each experiment three times, taking the median running time, among other
statistics, as measurements. The full experimental statistics can be found in the full version.

\begin{figure*}[t!]\captionsetup[subfigure]{font=scriptsize}
{
\begin{subfigure}[t]{0.49\textwidth}
    \centering\includegraphics[width=\columnwidth]{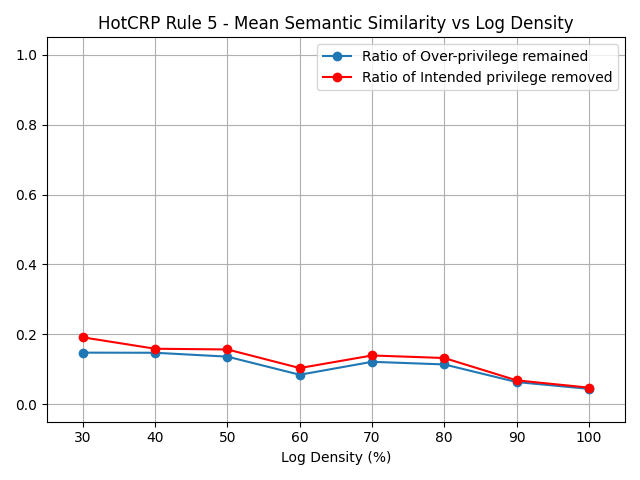}

    \caption{Semantic Similarity of the synthesized rule as a function of the log density, in terms of the over-privilege remained and the intended privilege removed (lower is better).}
    \label{fig:semantic-similarity}
\end{subfigure}
    \begin{subfigure}[t]{0.49\textwidth}
    \centering\includegraphics[width=\columnwidth]{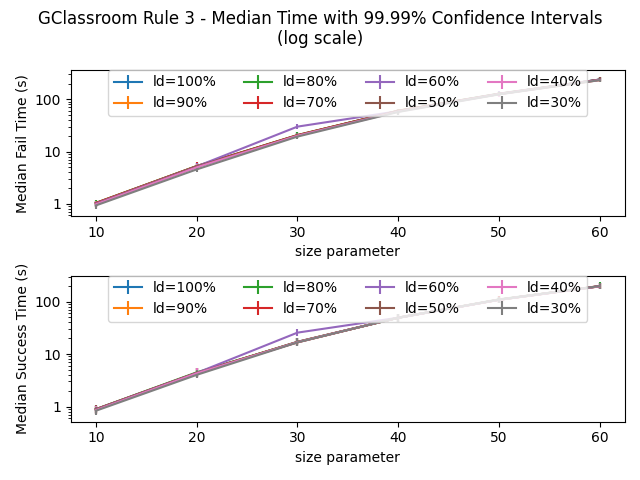}
    \caption{Running time (log scale) as a function of the size parameter and the log density (ld) with the 99.99\% confidence intervals shown as error bars.}
    \label{fig:median-time}
\end{subfigure}
}
\caption{Excerpt of the evaluation results in terms of semantic similarity of the generated rules versus the ideal ones and the scalability.}
\end{figure*}

\mypara{Q1: Effectiveness.}
We run \system{} on the input policy and compare its output to
$\policy_{\mathsf{tight}}$.
For cases where we do not get the intended tightening, we can still compute semantic similarities by enumerating all the over-privileges and intended privileges for each test case, and computing the percentage of over-privileges remaining and intended (but unexercised) privileges removed.
In the Google Classroom case study, all loosened rules are
tightened with the expected conjunct (except with added type guards).
For the HotCRP case study, two out of the five loosened rules were tightened to the ideal tight
rules, while the other three still achieved high semantic similarity. Figure~\ref{fig:semantic-similarity} shows the semantic similarity from one rule (with \texttt{size}$=30$) of the HotCRP case study, and that both semantic similarity metrics gets better (lower is better) as the log density increase, most likely because the real over-privileges got chosen instead of intended privileges in the request selection stage.
Our results also confirm that our 
approach of adding one conjunct at a time discourages
point solutions where we end up with an expression that denies one specific
point.

In short, in our case studies, \system{} successfully tightens most of the loose
rules, and preserves most rules that were not loosened.
On the few loosened rules where it was not
tightened ideally,
it still
removes some over-privileges while keeping most intended privileges, even when
the log density is low.

\mypara{Q2: Scalability.} To evaluate scalability, we measure \system's running time. We vary the problem size in two dimensions: the \texttt{size} parameter and the log density.
Figure~\ref{fig:median-time} shows the time for each successful and failed tightening from SyGuS when varying the \texttt{size} parameter and the log density (ld).
Figure~\ref{fig:median-time} is representative, showing that in general, \system
runs in sub-exponential time when increasing the number of entities
while being generally unaffected by the size of the log supplied into \system.
This shows that \system{} can handle problems of larger sizes efficiently.

\section{Related Work}

Manual development of\full{ effective and manageable} 
access control policies is well known to be challenging\full{ and error-prone}.
There is a sizable body of literature on algorithms for mining (\emph{a.k.a.}
learning\full{, synthesizing}) access control policies.\full{  Early work in this area
generated role-based policies from access control lists.}  Work on learning ABAC or ReBAC
policies starts with Xu et al.~\cite{xu14miningABAClogs,xu15miningABAC}
and now spans many
papers, e.g.,~\cite{mocanu2015towards,gautam2017constrained,talukdar2017efficient,bui2017mining,cotrini2018mining,bui19mining,cotrini2019next}. 
The general problem, like the policy tightening problem and for similar reasons, is under-constrained.
None of these algorithms is based on SyGuS.

There are only a few papers on algorithmically modifying or updating a
IAM-PolicyRefiner~\cite{dantoni2024automatically} and Eiers \etal~\cite{eiers2023quantitative} both tightens policies in
Amazon Web Services IAM (Identity and Access Management) policy language, and Mitani et al.~\cite{mitani2023qualitative} refines ABAC policies using machine learning.
IAM-PolicyRefiner and Eiers \etal, like \system, tightens one rule at a time. However, both restrict the
changes to a rule to \textit{local} changes to predicates that appear in the original rule.
\system is able to introduce predicates absent in the original policy. This makes tightening possible in more cases. The main focus of the two works are to generate regular expressions that restricts the name of the principal and resource, which is the main feature of IAM policies. We target Cedar, which is more expressive. Mitani \etal relies on ``qualitative intention'' as extra input, which \system does not. This makes Mitani \etal less suitable for cloud access control providers where the intention is not known.
\section{Discussion}
\label{sect:discussion}

\mypara{Assumptions and Limitations.} 
\system assumes the input policy is type-safe, and that the schema and environment are fixed. 
A limitation of \system is that the only considers tightening by adding a conjunction of atomic predicates to a rule.
Extending \system to consider more complex changes, such as adding arbitrary Boolean combinations of atomic predicates, is future
work.\full{ There are challenges, such as the fact that enumerating disjunctions can
cause an exponential blow-up in the number of predicates, especially with set
membership with constant sets. We confirmed the problem experimentally using our
case study setup, crafting ideal rules with disjunctions and set
memberships with constants. Some rules would apply to many types of principals,
actions, and resources. For example, we can condense the whole policy consisting
of multiple permit and deny rules into one complicated permit rule which is
difficult to tighten.
This can be addressed by, for example, manually guiding \system{} by
decomposing a rule into multiple disjoint rules applying to different actions.
An advantage of enumerating without disjunctions is that we do not generate
disjunctions of point solutions (\ie, predicates that exclude only a single
request).} 
We also assume that all permitted requests in the log should be permitted in the tightened policy. However, this is not always true in practice. This is because there might be accidental accesses in the log that exercise permissions that are not intended, or malicious parties identifying over-privileges themselves for exploitation. It is straightforward to extend \system to consider this case by loosening our semantic constraints to allow some of the entries in the log to not be permitted in the tightened rule. The formulation and experimental results can be found in the full version.

\mypara{Termination Criterion.}
By default, \system stops tightening a permit rule when SyGuS fails a specified number of times. \system{}  supports other termination criteria, such as a preset timeout or  
termination when a
certain percentage of tightening has been achieved, \ie, a certain percentage of $\pred{POP}$ has been removed.
Exploring such termination conditions is future work. 

\mypara{Rule overlap.} If there is overlap between rules (\ie, a request is
permitted because of two or more permit rules), then \system{}'s rule-level
tightening may not produce a policy that is as tight as expected.
This is because, in order to remove this overlap, \system would need to
successfully tighten all such rules, and each such tightening must exclude the
overlap.
This difficulty gives rise to an interesting, orthogonal direction of
future work: can the techniques employed by \system can be effectively adapted
to reducing overlap between rules, as an aid for policy maintainability?

\mypara{Manual Vetting.} Our intention with \system is for it to propose
tightened rules for vetting by administrators to ensure that they do not remove intended privileges.
Manual vetting is practical, as \system preserves the original rule’s
readability by limiting changes to each rule to adding conjuncts\full{, limiting the
complexity of candidate conjuncts, and never synthesizing new rules.
From our own experience with the case studies, \system often finds
\emph{exactly} the intended tightenings (modulo guards for type-safety), which
is a somewhat favorable indicator of readability}.

\mypara{Using SyGuS.}
Using a SyGuS solver to choose from a set of pre-computed predicates, as \system
currently does, is one of many methods to consider for generating and evaluating
potential tightenings\full{ for policy rules}.
Although we do not currently use SyGuS for predicate generation, we wish to keep this open for further extension, since a
bespoke term enumerator will likely be more difficult to extend.\full{
Recently, large language models (LLMs) have been applied to code generation
tasks~\cite{10772760}, and so provide another interesting avenue for further exploration.}

Concerning the evaluation stage, one can encode the predicate directly
into SMT constraints, but this would also require encoding \emph{type
  safety} constraints.
As we discussed in Section~\ref{sect:symbolic-compilation}, our approach 
avoids this difficulty by providing the SyGuS solver with \emph{type-safe
  predicates}, effectively enforcing type safety through the grammar.

\mypara{Missing features of Cedar.} 
Currently, there are features in Cedar that we could not generate predicates for tightening. However, \system can be modularly extended, under some assumptions, to support them. For example, predicates of the form $x \in S$ where $S$ is a constant set can be generated by limiting the cardinality of $S$ to constants that appeared in the log and entity store. One challenging aspect is synthesizing linear integer arithmetic (LIA) predicates of the form x + y < c, where c is a symbolic constant to be synthesized, though there are existing approaches to this problem~\cite{abate2023synthesising}.
\bibliographystyle{splncs04}
\bibliography{ref}

\end{document}